\documentclass[aps,prd,reprint]{revtex4-1}
\usepackage{graphicx}
\usepackage{enumerate}
\usepackage{hyperref}
\usepackage{textcomp}
\usepackage{amsmath,amsfonts,amstext}
\usepackage{pgf}
\usepackage[utf8]{inputenc}
\usepackage{braket}
\usepackage{mathtools}
\usepackage{indentfirst}
\usepackage{epsfig}
\usepackage{subcaption}
\usepackage{hyperref}
\usepackage{xcolor}
\allowdisplaybreaks
\begin{document}

	\title{ 
	Constant-roll warm inflation and the $\beta$-function approach}
		\author{ Ui Ri Mun}
	\affiliation{
		Department of Physics, Kim Il Sung University, \\ Ryongnam Dong, Taesong District, Pyongyang, Democratic People's Republic of Korea
	}

	\begin{abstract}
		\noindent 
		We propose a new approach to constant-roll warm inflation as a generalization of constant-roll inflation. Based on the $\beta$-function formalism, it is shown that constant-roll warm inflation models with a natural end fall into universality classes defined by three different types of $\beta$-functions, under the assumption that radiation energy density is quasi-stable. Given that warm inflation is completely specified by the $\beta$-function and dissipation coefficient ratio, we investigate whether or not the inflation can physically be realized for enough number of e-foldings at the background level for some combinations of the $\beta$-functions and non-trivial dissipation coefficient ratios. 
	\end{abstract}
		
	\maketitle
	
	\section{Introduction}
One of the cornerstones of modern cosmology is cosmic inflation \cite{Guth,Linde,Linde1,starobinsky,steinhardt}, i.e., the accelerated expansion of the early universe. Inflation provides an elegant solution to the puzzles in the standard Big Bang cosmology such as the horizon and flatness problems, and moreover, well explains the origin of anisotropies observed in the cosmic microwave background (CMB) radiation \cite{inhom1,inhom2,inhom3}. The simplest and common candidate which can drive inflation is a scalar field, called the inflaton,  for obvious reasons. To have enough number of e-foldings during inflation, the inflaton potential is normally chosen to be flat enough so that the slow-roll conditions are satisfied. There have been known a myriad of inflation models that respect the slow-roll conditions and observational constraints by Planck Collaboration \cite{Planck}. An alternative to slow-roll inflation is constant-roll inflation \cite{cr-2}, which requires the second slow-roll parameter to be constant that is not necessarily small. Constant-roll inflation has received much attention these days, as it can not only produce a nearly scale invariant power spectrum of scalar perturbations consistent with observations \cite{cr-obs-1,cr-obs-2} but also predict non-Gaussianities \cite{NG-1,NG-2,NG-3}.

Although the recent Planck results \cite{Planck}  excluded several inflaton potentials, they are still not accurate enough to pin down a concrete potential. Moreover, from a purely theoretical point of view, there is no preferred inflaton potential. Using potential, therefore, would not be the best way to classify inflation models in view of current status of observational and theoretical cosmology. 
In Ref. \cite{beta-cold}, the authors proposed  the $\beta$-function formalism of inflation, based on a formal analogy between the renormalization group equation (RGE) in quantum field theory (QFT) and the equation for the evolution of the inflaton field. In fact, $\beta$-function formalism is theoretically supported by the holography theory applied to cosmology (see e.g. \cite{hol-1,hol-2,hol-3}). In this formalism, inflation is interpreted as a flow of the inflaton field away from the fixed point that corresponds to the exact de Sitter geometry. Therefore, the characterization of $\beta$-function near a fixed point can define a universality class of inflation models with discriminative potentials. The $\beta$-function formalism was indeed successfully applied not only to  slow-roll inflation \cite{beta-sr-1,beta-sr-2} but also to constant-roll inflation \cite{Francesco}. The formalism was extended in Ref. \cite{berera} to warm inflationary paradigm \cite{WI-1,WI-2,WI-3}.

In this work, building on the results of Refs. \cite{Francesco,berera}, we propose a new approach to constant-roll warm inflation which generalizes the ordinary constant-roll inflation studied in Ref. \cite{Francesco}. By employing the $\beta$-function formalism, we show that the constant-roll warm inflationary condition can be rendered into an equation for $\beta$-function that has analytic solutions in the presence of non-trivial dissipation. We investigate whether or not each $\beta$-function satisfying constant-roll condition is suitable to describe an inflation model with a natural end for non-trivial dissipation coefficient ratios. We also find some general parameterizations of the dissipation coefficient ratios compatible with the constant-roll warm inflationary condition. 

This article is organized as follows. In Section \ref{sec-1}, we briefly review the $\beta$-function formalism of single field warm inflation and classify possible behaviors of $\beta$-function to which the corresponding inflation model can naturally end. Section \ref{CWI} is the main part of our paper, in which we present constant-roll warm inflationary condition and find the $\beta$-functions satisfying it. We then proceed for given non-trivial dissipation coefficient ratios to determine which $\beta$-function can indeed be used to describe the constant-roll warm inflation. We conclude in Section \ref{sec-3}.

	\section{$\beta$-function formalism of single field warm inflation}\label{sec-1}
	In single field warm inflation models, the dynamics of the background inflaton field $\phi$ minimally coupled to gravity, in a flat Friedmann-Lema\^itre-Robertson-Walker (FLRW) universe, is  determined by the following set of equations
	\begin{align}
		&\ddot{\phi}+3H(1+Q)\dot{\phi}+\frac{dV}{d\phi}=0\,,\label{EOM-phi}\\
		&H^2=\frac{\rho_\phi+\rho_r}{3}\,,\label{Hubble-1}\\
		&\dot{H}=-\frac{1}{2}(\rho_\phi+P_\phi+\rho_r+P_r)=-\frac{1}{2}\left(\dot{\phi}^2+\frac{4}{3}\rho_r\right)\,,\label{Hubble-2}\\
		&\dot{\rho}_r+4H\rho_r=\Upsilon\dot{\phi}^2\,,\label{rho-r}\\
		&\rho_\phi=\frac{1}{2}\dot{\phi}^2+V(\phi)\,,\quad P_\phi=\frac{1}{2}\dot{\phi}^2-V(\phi)\,.\label{rho-p-phi}
	\end{align} 
These equations were written in natural units $M_P^2\equiv (8\pi G)^{-1}=1$ and as usual, overdots indicate  time derivative. $H=\dot{a}/a$ is the Hubble parameter with scale factor $a(t)$, $\rho$ and $P$ are the energy density and  pressure with the subscripts $\phi,\,r$ used for the inflaton field and for the radiation, respectively.
	 $V(\phi)$ denotes the effective inflaton potential and $Q\equiv \Upsilon/3H$ the dissipation coefficient ratio with the dissipation coefficient $\Upsilon$ that depends on the interaction between the inflaton and other particles, e.g. standard model (SM) particles, in radiation bath.   The equations \eqref{EOM-phi}-\eqref{rho-p-phi} can  be solved   if the effective potential $V(\phi)$ and the dissipation coefficient ratio $Q$ are known.	
	The dissipation coefficient ratio is in general parameterized as
	\begin{equation}
		Q(T,\phi)=\frac{CT^r}{H\phi^s}\,,\label{Q-general}
	\end{equation}
	where $T$ is the temperature of the radiation bath to which the inflaton field is coupled and $C$ is a constant (see e.g. Refs. \cite{Q-1,Q-2,Q-3,Q-4,Q-5,Q-6,Q-7,Q-8}). Alternatively, one can  view the dissipation coefficient as a function of $\phi$, i.e., $Q(\phi)$ and then find $T(\phi)$ and the corresponding parameterization given by eq. \eqref{Q-general}, as suggested in Ref. \cite{berera}. We will adopt this viewpoint in the paper. We also assume that the radiation energy density is quasi-stable, i.e., 
	\begin{equation}
\left\vert\frac{\dot{\rho}_r}{4H\rho_r}\right\vert\ll 1\,.\label{qasi-con}
	\end{equation}

	The $\beta$- function is defined as
	\begin{equation}\label{beta-defn}
	\beta(\phi)=\frac{d \phi}{d\ln a}\equiv \frac{d\phi}{dN}=\frac{\dot{\phi}}{H}\,,
	\end{equation} 
and it can be used in place of the inflaton potential to classify both cold and warm inflation models with some advantages in the study of universality classes of inflation models. The notation $\beta(\phi)$ attributes to the close resemblance to the well known $\beta$-function in QFT. Here $\phi$ plays the role of the coupling constant and the number of e-foldings $N$ the role of the renormalization scale.

Assume that $\phi(t)$ is monotonic at least during inflation so that its inverse $t(\phi)$ can be found. Let  $W(\phi)\equiv -2H(\phi)$ be the superpotential and use the assumption of quasi-stable radiation energy density in eqs. \eqref{Hubble-2} and \eqref{rho-r} to get 
\begin{equation}
\dot{\phi}\simeq\frac{W_{,\phi}}{1+Q}\label{phi-dot}\,,
\end{equation}
which combines with eq. \eqref{beta-defn} to give
\begin{equation}
(1+Q)\beta\simeq\frac{-2W_{,\phi}}{W}\,.\label{beta-CI}
\end{equation}
Here $W_{,\phi}\equiv \frac{dW}{d\phi}$ and hereafter we will use such notation to indicate the derivative with respect to $\phi$ for notational simplicity. In Ref. \cite{berera}, the authors introduced the quantity $\beta_\textrm{CI}\equiv \beta(1+Q)$ to classify  warm inflation models. However, we will keep making use of $\beta$. The reason will be clear in Section \ref{CWI}.
Once the $\beta$-function is known, one can compute the superpotential using eq. \eqref{beta-CI} as 
\begin{equation}
W(\phi)\simeq W_f\exp\left[-\frac{1}{2}\int_{\phi_f}^{\phi}(1+Q(\phi'))\beta(\phi')\,d\phi'\right]\,,\label{superpotential}
\end{equation}
where we used the subscript $``f"$ to indicate quantities evaluated at the end of inflation.
We can express the inflaton potential  by using the $\beta$-function with the help of eq. \eqref{Hubble-1}
along with eqs.  \eqref{phi-dot} and \eqref{beta-CI} as
\begin{equation}
V(\phi)\simeq\frac{3}{4}W^2(\phi)\left[1-\frac{1}{6}\left(1+\frac{3}{2}Q(\phi)\right)\beta^2(\phi)\right]\,.\label{potential-beta}
\end{equation}
It is also useful to write $\rho_r$ and $\rho_\phi$ in terms of $\beta$-function. It follows from eqs. \eqref{Hubble-1}, \eqref{rho-r} and \eqref{beta-defn} that
\begin{align}
	\rho_r&=\frac{3}{16}\beta^2W^2Q\,,\label{betaQ-rho-r}\\
\rho_\phi&=\frac{3}{16}W^2(4-\beta^2Q)\,.\label{betaQ-rho-phi}
\end{align}
 From eqs. \eqref{betaQ-rho-r} and \eqref{betaQ-rho-phi}, we see that the ratio between the inflaton energy density and the radiation energy density is completely determined by $\beta^2Q$, i.e.,
\begin{equation}
\frac{\rho_r}{\rho_\phi}=\frac{\beta^2Q}{4-\beta^2Q}\,,\label{rho_r-rho_phi}
\end{equation}
which shows explicitly that $\rho_r/\rho_\phi$ increases with $\beta^2Q$. 
Assuming that the radiation is thermalized within Hubble time, we have
\begin{equation}
\rho_r=\frac{\pi^2 g_*}{30}T^4\label{rho-r-T}\,,
\end{equation}
with $g_*$ being  the number of relativistic degrees of freedom at temperature $T$ for the radiation bath. The quantity $\beta^2Q$ is related to $\beta_T$ introduced  in Ref. \cite{berera} by $\beta^2 Q=\frac{2\pi^2 g_*}{45} T^2\beta_T^2$.  Using eqs. \eqref{betaQ-rho-r} and  \eqref{rho-r-T}, the temperature of the radiation bath is obtained as 
\begin{equation}
	T=\left(\frac{45}{8\pi^2g_*}\beta^2W^2Q\right)^{1/4}\,.
\end{equation}
 To characterize the inflationary phase, let us express the equation of state in terms of $\beta$-function. Using eqs. \eqref{Hubble-2}, \eqref{beta-defn} and \eqref{phi-dot}, we find
\begin{equation}\label{equation-of-state}
1+w\equiv\frac{-2\dot{H}}{3H^2}\simeq\frac{1+Q}{3}\beta^2\,,
\end{equation}
which implies that  inflation can be realized when $\vert\beta\vert\sqrt{1+Q}<\sqrt{2}$ and the exact de Sitter corresponds to $\beta=0$. Throughout the above discussions, we see that the $\beta$-function along with the dissipation coefficient ratio allows us to have full understanding of warm inflation dynamics at the background level.

\begin{figure}[h!]	
	\centering
	\begin{subfigure}{0.2\textwidth}
		\includegraphics[scale=0.4]{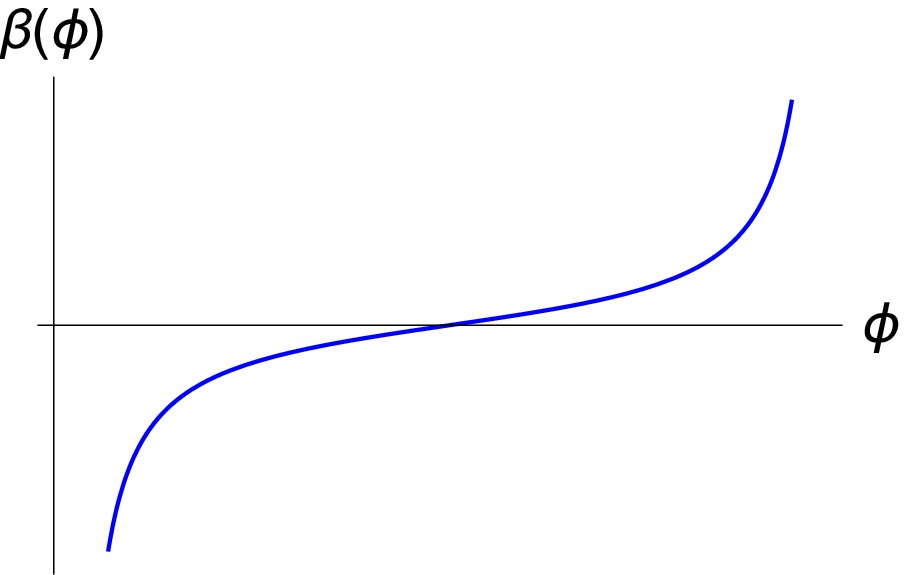}
		\subcaption{}\label{beta-type-a}
	\end{subfigure}
	\begin{subfigure}{0.2\textwidth}
		\includegraphics[scale=0.4]{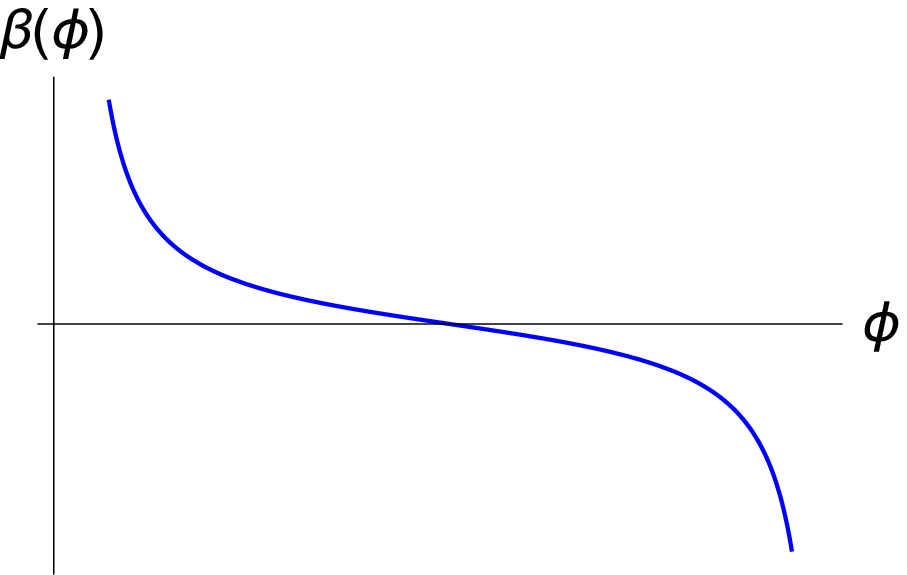}
		\subcaption{}\label{beta-type-b}
	\end{subfigure}
	\begin{subfigure}{0.2\textwidth}
		\includegraphics[scale=0.4]{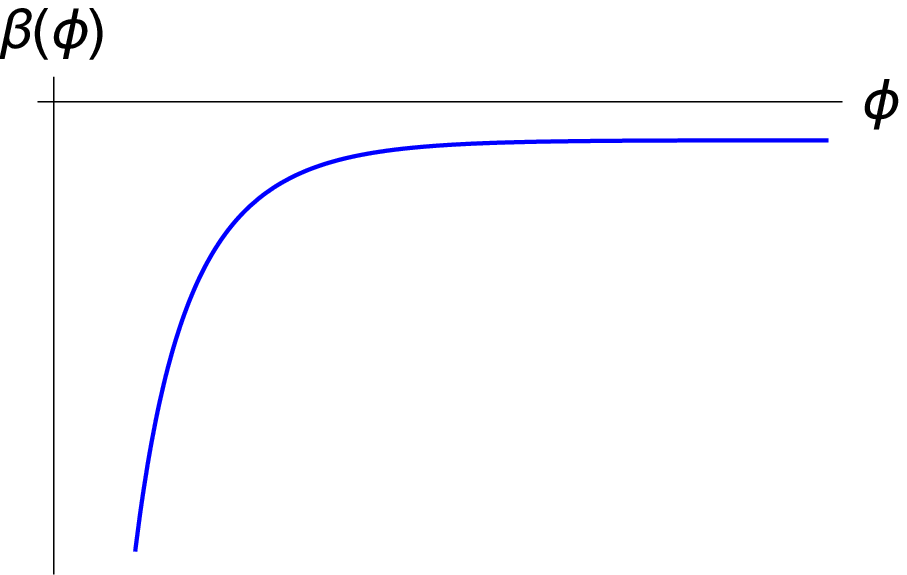}
		\subcaption{}\label{beta-type-c}
	\end{subfigure}
	\begin{subfigure}{0.2\textwidth}
		\includegraphics[scale=0.4]{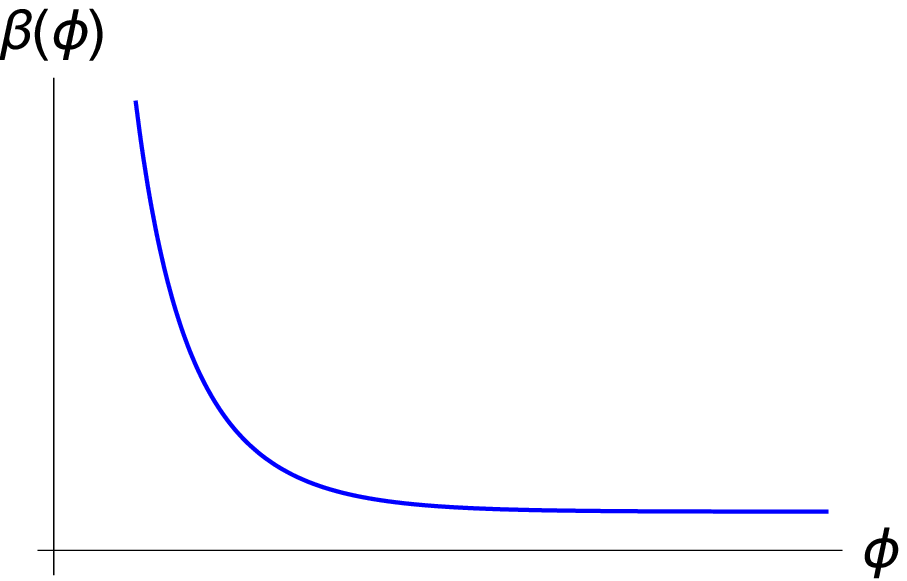}
		\subcaption{}\label{beta-type-d}
	\end{subfigure}
	\caption{Possible behaviors of $\beta(\phi)$ near a fixed point. The curves do not correspond to a $\beta$-function of a certain inflation model and are drawn for illustrative purposes only. }\label{beta-type}
\end{figure}

In analogy with renormalization group (RG) approach in QFT, we identify the zeros of $\beta(\phi)$ as fixed points. Here we emphasize that the natural end of inflation can only be attained if $\phi$ moves away from the fixed point as the universe expands, namely, the number of e-foldings increases. This, in the language of RG approach, implies that the $\beta$-function must have an \textit{infrared} fixed point, which requires  $\beta(\phi)$ to be a strictly increasing function of $\phi$ in the neighborhood of the fixed point, that is, the first derivative $\beta_{,\phi}=\frac{d\beta(\phi)}{d\phi}$ should be positive. Indeed, this can be illustrated in Fig. \ref{beta-type}. 
$\beta(\phi)$ depicted in Fig. \ref{beta-type-a} increases with $\phi$. Since $\beta(\phi)>0$ for $\phi$ larger than the value corresponding to the fixed point and $\beta(\phi)<0$ otherwise, by definition of $\beta$-function, we see that $\phi$ moves away from the fixed point as the universe expands. On the contrary, for $\beta(\phi)$ depicted in Fig. \ref{beta-type-b} that decreases with $\phi$, $\phi$ moves towards the fixed point as the universe expands. The $\beta$-functions shown in Figs. \ref{beta-type-c} and \ref{beta-type-d} do not have a fixed point, instead they approach a constant value when $\phi$ goes to infinity. Such $\beta$-functions can  also describe the inflation models, as it was mentioned in Ref.  \cite{beta-cold}, as long as the constant value that the $\beta$-function approaches is small enough.  For warm inflation models, the quantity $\beta\sqrt{1+Q}$ rather than $\beta$ itself has to approach a small value, as it can be seen from eq. \eqref{equation-of-state}. The $\beta$-function in Fig. \ref{beta-type-c} increases with $\phi$, hence  $\phi$ moves away from a constant value and the corresponding inflation model can end naturally, while the $\beta$-function in Fig. \ref{beta-type-d} that decreases with $\phi$ can not describe an inflation model with a natural end, for the same reason as we have mentioned above.\\

\section{Constant-roll warm inflation}\label{CWI}
\subsection{Universality classes}
\textit{Constant-roll warm inflation} is specified by the relation
\begin{equation}\label{const-roll-1}
\ddot{\phi}=-3H(1+Q)\dot{\phi}\lambda
\end{equation}
with constant parameter $\lambda$. In terms of the second slow-roll parameter $\eta\equiv -\ddot{H}/(2H\dot{H})$, the above condition can equivalently written as 
\begin{equation}
\eta/(1+Q)=3\lambda\,. \label{const-roll-eta}
\end{equation}
Notice the difference between the constant-roll condition given by eq. \eqref{const-roll-1} (or eq. \eqref{const-roll-eta}) and $\ddot{\phi}=-3H\dot{\phi}\lambda$ (or equivalently, $\eta=\textrm{const}$) discussed in Ref. \cite{const-roll-warm}. 

In what follows we will find the $\beta$-function which is consistent with the constant-roll warm inflationary condition \eqref{const-roll-1} under the assumption of quasi-stable radiation energy density. 
 Take the time derivative of eq. \eqref{phi-dot} and use the chain rules to obtain
 \begin{equation}
 	\ddot{\phi}\simeq\dot{\phi}\left[\frac{W_{,\phi\phi}}{1+Q}-W_{,\phi}\frac{Q_{,\phi}}{(1+Q)^2}\right]\,.\label{phi-ddot}
 \end{equation}
Eq. \eqref{beta-CI} allows us to rewrite eq. \eqref{phi-ddot} as
\begin{equation}\label{phi-ddot-2}
	\ddot{\phi}\simeq\frac{\dot{\phi}W}{2}\left[(1+Q)\frac{\beta^2}{2}-\beta_{,\phi}\right]\,.
\end{equation}
Plugging eq. \eqref{phi-ddot-2} into eq. \eqref{const-roll-1}, we get
\begin{equation}
	\frac{\beta^2}{2}-\frac{\beta_{,\phi}}{1+Q}\simeq 3\lambda\,.\label{const-roll-2}
\end{equation}
If we used $\beta_\textrm{CI}$ instead of $\beta$, the constant-roll condition would not be translated into a simple equation as eq. \eqref{const-roll-2}. 
The solutions to eq. \eqref{const-roll-2} can be found as
\begin{widetext}
\begin{equation}\label{beta-sol}
	\beta(\phi)=\left\lbrace\begin{array}{cc}
\sqrt{6\lambda}\,\coth\left[\sqrt{\frac{3\lambda}{2}}\int_{\phi}^{0}(1+Q(\phi'))d\phi'\right]&\text{for $\lambda>0$ and $\beta>\sqrt{6\lambda}$}\\
\sqrt{6\lambda}\,\tanh\left[\sqrt{\frac{3\lambda}{2}}\int_{\phi}^{0}(1+Q(\phi'))d\phi'\right]&\text{for $\lambda>0$ and $\beta<\sqrt{6\lambda}$}\\
\sqrt{6\lvert\lambda\rvert}\,\cot\left[\sqrt{\left\vert\frac{3\lambda}{2}\right\vert}\int_{\phi}^{0}(1+Q(\phi'))d\phi'\right]&\text{for $\lambda<0$}\\
\sqrt{6\lvert\lambda\rvert}\,\tan\left[-\sqrt{\left\vert\frac{3\lambda}{2}\right\vert}\int_{\phi}^{0}(1+Q(\phi'))d\phi'\right]&\text{for $\lambda<0$}
	\end{array}\right.\,.
\end{equation}
\end{widetext}
When there is no dissipation in the inflaton energy , i.e., $Q\to 0$, it is obvious that the above solutions recover those obtained in \cite{Francesco}, as they should be. Note that not all solutions in eq. \eqref{beta-sol} are suitable to describe the inflation models with a natural end. To see this, let us rewrite eq. \eqref{const-roll-2} as 
\begin{equation}
	\beta_{,\phi}\simeq\left(\frac{\beta^2}{2}-3\lambda\right)\left(1+Q\right)\,.\label{beta-der}
\end{equation}
Since $1+Q\geq 1$, eq. \eqref{beta-der} implies  $\beta_{,\phi}>0$ for the solutions except the second solution in eq. \eqref{beta-sol}. We thus conclude that constant-roll warm inflation models with a natural end fall into universality classes defined by three different types of $\beta$-functions. 

At this point, it is fair to emphasize that all solutions given in eq. \eqref{beta-sol} must obey the assumption of quasi-stable radiation energy density given by \eqref{qasi-con}, which can written in terms of $\beta$-function as (cf. \cite{berera})
\begin{equation}
\left\vert\frac{\beta_{,\phi}}{2}-\frac{\beta^2}{4}(1+Q)+\frac{\beta}{4}\frac{Q_{,\phi}}{Q}\right\vert\,\ll 1.
\end{equation}
Using eq. \eqref{beta-der}, the above relation boils down to 
\begin{equation}
	\left\vert\frac{\beta}{4}\frac{Q_{,\phi}}{Q}-\frac{3\lambda}{2}(1+Q)\right\vert\ll 1\,.\label{qs-con-c}
\end{equation}
We kindly refer readers to Ref. \cite{const-roll-warm} for an analytic solution to constant-roll warm inflation with $Q=\textrm{const}$ that is not constrained by the assumption of quasi-stable radiation energy density.

In the following sections, we investigate if the $\beta$-functions given by eq. \eqref{beta-sol} are  suitable to realize inflationary phase with sufficient number of e-foldings (e.g. $N\geq 50$) for some toy models of the dissipation coefficient ratios.\\

\subsection{Hyperbolic $\beta$-function}
We begin with the first solution of $\beta(\phi)$ in eq. \eqref{beta-sol}, for which the quantity $\beta(\phi)\sqrt{1+Q(\phi)}$ could be well below than $\sqrt{2}$ for sufficiently small $\lambda$ so that the inflationary phase can be realized.  Also, the derivative of the $\beta(\phi)$ with respect to $\phi$ is 
\begin{widetext}
\begin{equation}
	\beta_{,\phi}=\sqrt{6\lambda}\,\textrm{csch}^2\left[-\sqrt{\frac{3\lambda}{2}}\int_{0}^{\phi}(1+Q(\phi'))d\phi'\right]\left[\sqrt{\frac{3\lambda}{2}}(1+Q(\phi))\right]\,,
\end{equation}
which is indeed positive given that $Q(\phi)\geq 0$. Hence, the corresponding inflation model to this $\beta$-function can naturally end.

The superpotential is obtained by using eq. \eqref{superpotential} as follows:
\begin{align}
	W(\phi)&\simeq W_f\exp\left[-\frac{\sqrt{6\lambda}}{2}\int_{\phi_f}^{\phi}(1+Q(\phi'))\coth\left(\sqrt{\frac{3\lambda}{2}}\int_{\phi'}^{0}(1+Q(\phi''))d\phi''\right)\,d\phi'\right]\nonumber\\
	&=W_f\exp\left[-\frac{\sqrt{6\lambda}}{2}\int_{\phi_f}^{\phi}\coth\left(\sqrt{\frac{3\lambda}{2}}\int_{\phi'}^{0}(1+Q(\phi''))d\phi''\right)\,d\left(\int_{0}^{\phi'}(1+Q(\phi''))d\phi''\right)\right]\nonumber\\
	&=W_f\exp\left[\ln\left(\frac{\sinh\left(-\sqrt{\frac{3\lambda}{2}}\int_{0}^{\phi}(1+Q(\phi'))d\phi'\right)}{\sinh\left(-\sqrt{\frac{3\lambda}{2}}\int_{0}^{\phi_f}(1+Q(\phi'))d\phi'\right)}\right)\right]\nonumber\\
	&={W}_f\frac{\sinh\left[-\sqrt{\frac{3\lambda}{2}}\int_{0}^{\phi}(1+Q(\phi'))d\phi'\right]}{\sinh\left[-\sqrt{\frac{3\lambda}{2}}\int_{0}^{\phi_f}(1+Q(\phi'))d\phi'\right]}\nonumber\\
	&\equiv\overline{W}_f\sinh\left[-\sqrt{\frac{3\lambda}{2}}\int_{0}^{\phi}(1+Q(\phi'))d\phi'\right]\,.\label{superpotential-sol}
\end{align}
\end{widetext}
Here  $\overline{W}_f$ is the normalization constant to be determined by observational data and the second equality is followed by the fundamental theorem in  calculus,
\begin{equation}
\frac{d}{dx}\int_{x_0}^x f(x')dx'=f(x)\,.
\end{equation}
Accordingly, the Hubble parameter is given by
\begin{equation}
	{H(\phi)\simeq -\frac{1}{2}\overline{W}_f\sinh\left[-\sqrt{\frac{3\lambda}{2}}\int_{0}^{\phi}(1+Q(\phi'))d\phi'\right]}\,.\label{Hubble-phi-1}
\end{equation}
On the other hand, using eq. \eqref{superpotential-sol} in eq. \eqref{potential-beta} gives
\begin{widetext}
\begin{align}
	V(\phi)&\simeq\frac{3}{4}\overline{W}_f^2\left[1-\frac{1}{6}\left(1+\frac{3}{2}Q(\phi)\right)\beta^2(\phi)\right]\sinh^2\left[-\sqrt{\frac{3\lambda}{2}}\int_{0}^{\phi}(1+Q(\phi'))d\phi'\right]\nonumber\\
	&=\frac{3}{4}\overline{W}_f^2\left\lbrace-\lambda\left(1+\frac{3}{2}Q(\phi)\right)+\left[1-\lambda\left(1+\frac{3}{2}Q(\phi)\right)\right]\sinh^2\left[-\sqrt{\frac{3\lambda}{2}}\int_{0}^{\phi}(1+Q(\phi'))d\phi'\right]\right\rbrace\,.\label{potential-sol}
\end{align}

The number of e-foldings during inflation is given by 
\begin{align}\label{N1}
	N=-\int_{\phi_f}^{\phi}\frac{d\phi'}{\beta(\phi')}\simeq\frac{1}{\sqrt{6\lambda}}\int_{\phi}^{\phi_f}\tanh\left[\sqrt{\frac{3\lambda}{2}}\int_{\phi'}^{0}(1+Q(\phi''))d\phi''\right]d\phi'\,.
\end{align}
\end{widetext}
In order to see if the $\beta$-function under consideration can describe an inflation model, we consider two simple cases for $Q(\phi)$ as follows. 
\begin{itemize}
	\item $Q=a \,\phi^{n}$, for $n\in \textbf{N}$ and $a\in\textbf{R}$.
	
	The $\beta$-function is obtained as 
	\begin{equation}\label{beta-11}
		\beta(\phi)=\sqrt{6\lambda}\coth\left[-\sqrt{\frac{3\lambda}{2}}\left(\phi+\frac{a\,\phi^{n+1}}{n+1}\right)\right]\,,
	\end{equation}
	which does not have a fixed point, but approaches a constant value when $\phi\to -\infty$. 	As we mentioned earlier, it is not necessary to have a fixed point to realize an inflationary phase. Cold inflation can be realized even if $\beta$-function approaches a small constant value, more precisely, a value smaller than $\sqrt{2}$. However, in the presence of dissipation in the inflaton energy,   to have an inflationary phase,  what we need is  $\vert\beta(\phi)\vert\sqrt{1+Q(\phi)}$ that has a fixed point or  approaches a constant value smaller than $\sqrt{2}$. For the dissipation coefficient ratio under consideration, none of these can  be attained, therefore the $\beta$-function given by eq. \eqref{beta-11} is inappropriate for inflation.

\item $Q=a\,\phi^{-n}$, for $n\in \textbf{N}$ and $a\in\textbf{R}$

The $\beta$-function becomes
	\begin{equation}\label{beta-12}
\beta(\phi)=\sqrt{6\lambda}\coth\left[-\sqrt{\frac{3\lambda}{2}}\left(\phi+\frac{a\,\phi^{1-n}}{1-n}\right)\right]\,.
\end{equation}
Again, the $\beta$-function does not have a fixed point and approaches a constant value when $\vert\phi\vert\to \infty$. Differently from the previous case, the quantity $\beta(\phi)\sqrt{1+Q(\phi)}$ has the same asymptotic behavior as $\beta(\phi)$, because $Q(\phi)$ goes to zero when $\phi\to \infty$.
Since $\vert\phi\vert$ decreases as the universe expands, one may think that the inflation can start in weak dissipative regime ($Q(\phi)\ll 1$) and end in strong dissipative regime ($Q(\phi)\gg 1$). This, however, is not true for $n=2$. Indeed, in order for inflation to happen in strong dissipative regime, we need $\beta(\phi)<1$ and $Q(\phi)>1$, which read for the case under consideration
\begin{align}
\beta(\phi)&=\sqrt{6\lambda}\coth\left[-\sqrt{\frac{3\lambda}{2}}\left(\phi+\frac{a\,\phi^{1-n}}{1-n}\right)\right]<1\label{beta-s-1}\,,\\
Q(\phi)&=a\phi^{-n}>1\,.\label{Q-g-1}
\end{align}
It is not difficult to show that the above  inequalities can not simultaneously be satisfied for $n=2$ and $\phi<0$. For $n> 2$, we find that for large absolute value of $a$, more precisely, $\vert a\vert\geq \left(\frac{2(n-1)}{n-2}\right)^n$, the inequalities \eqref{beta-s-1} and \eqref{Q-g-1} can simultaneously hold true. 
To visualize the energy transfer from inflaton field to radiation bath during inflation, we plot the ratio of  radiation energy density to inflaton energy density along with the corresponding dissipation coefficient ratio in Fig. \ref{rho-Q-1} for different values of parameters. 
\begin{figure}[htp!]
	\centering
	\begin{subfigure}{0.23\textwidth}
		\includegraphics[width=1\linewidth,height=3cm]{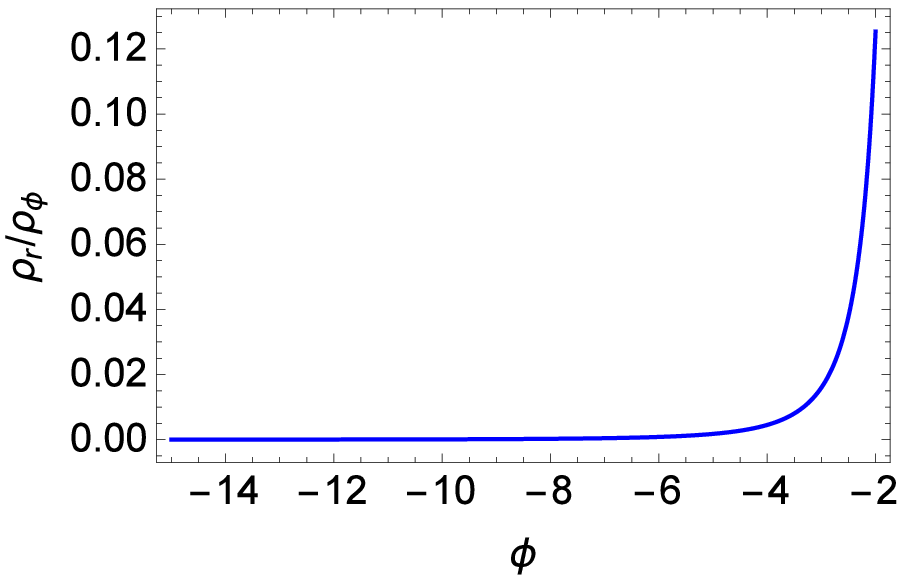}
	\end{subfigure}
\begin{subfigure}{0.23\textwidth}
	\includegraphics[width=1\linewidth,height=3cm]{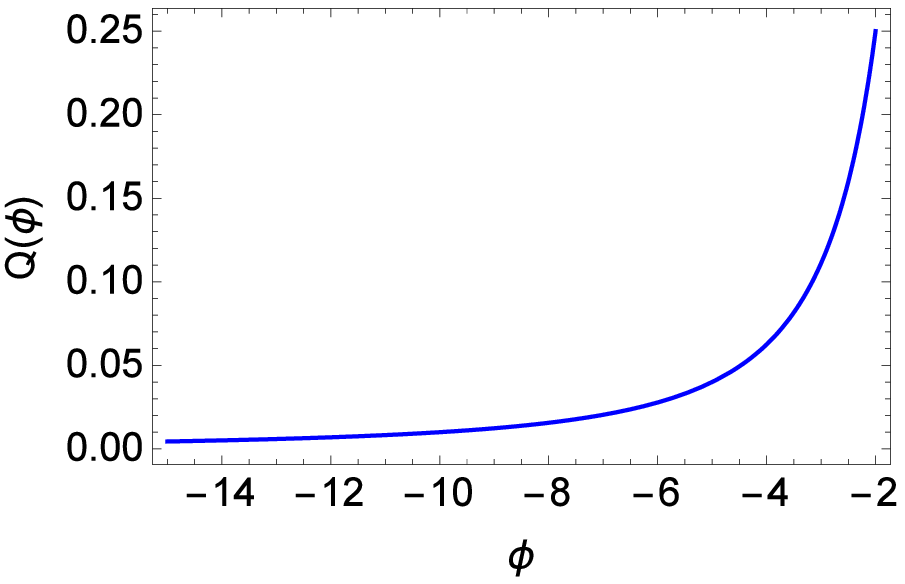}
\end{subfigure}
	\begin{subfigure}{0.23\textwidth}
		\includegraphics[width=1\linewidth,height=3cm]{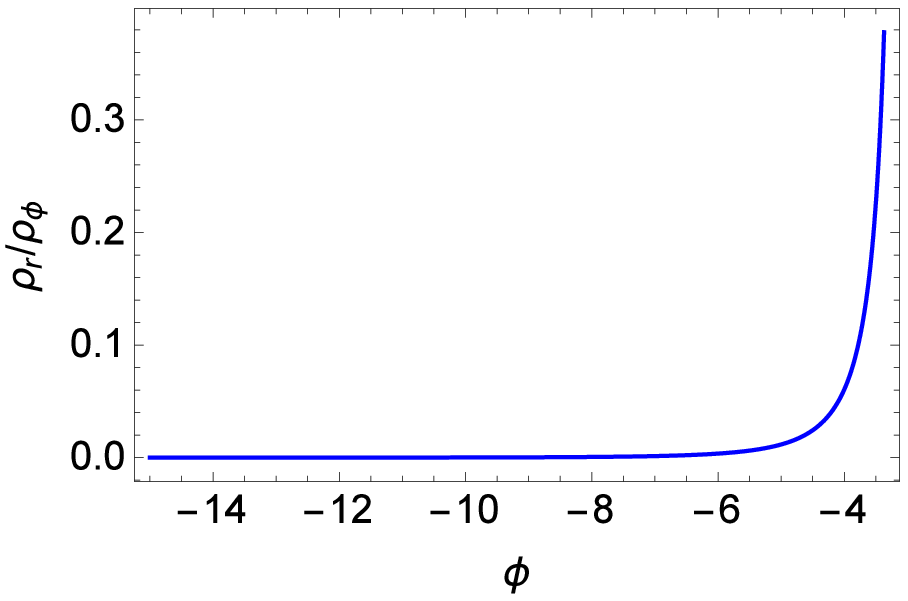}
	\end{subfigure}
\begin{subfigure}{0.23\textwidth}
	\includegraphics[width=1\linewidth,height=3cm]{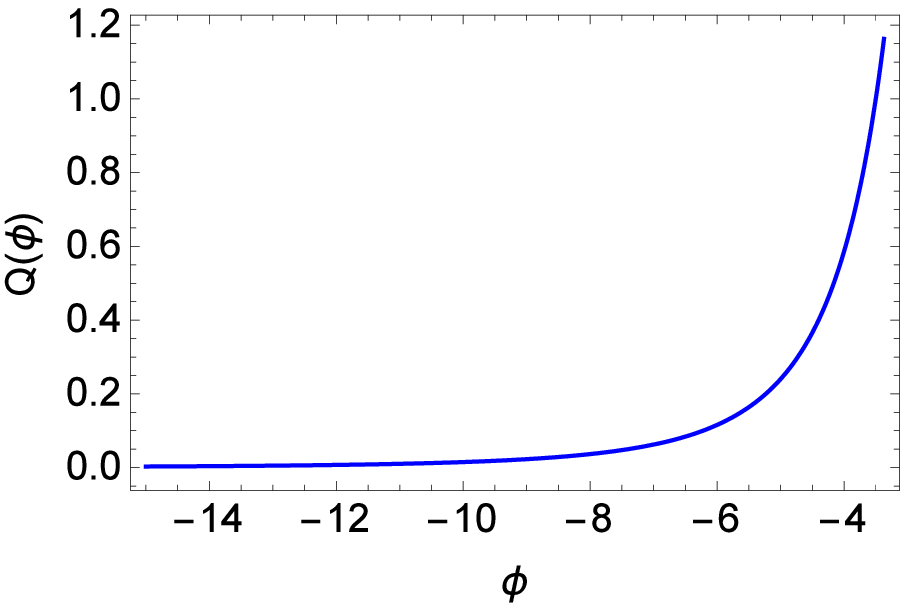}
\end{subfigure}
	\caption{The evolution of $\rho_r/\rho_\phi$ and $Q(\phi)=\phi^{-n}$ for $\beta(\phi)$ given by eq. \eqref{beta-12} with $\lambda=0.001$ during inflation. We have chosen $a=1$, $n=2$ (top) and $a=150$, $n=4$ (bottom). The range of $\phi$ in each plot was chosen such that the number of e-foldings satisfies $N\sim 50$. For $n=2$, $Q(\phi)\ll 1$ is satisfied during inflation and the corresponding model remains in weak dissipative regime. For $n=4$, $Q(\phi)$  increases monotonically from $Q(\phi)\ll 1$ to $Q(\phi)\gg 1$ and the corresponding model enters gradually into strong dissipative regime getting close to the end of inflation.}\label{rho-Q-1}
\end{figure}  
The temperature of the radiation bath can be found as
\begin{equation}
T(\phi)\simeq\sqrt{\left\vert\overline{W}_f\sqrt{6\lambda\,a\phi^{-n}}\,\cosh\left[-\sqrt{\frac{3\lambda}{2}}\left(\phi+\frac{a\,\phi^{1-n}}{1-n}\right)\right]\right\vert}\,.\label{T-phi-1}
\end{equation}  
We see from eqs. \eqref{Hubble-phi-1} and \eqref{T-phi-1} that the dissipation coefficient ratio $Q=a\phi^{-n}$ corresponds to $r=2$ and $s=n/2$ in the general parameterization in eq. \eqref{Q-general}, namely $Q(\phi,T)=\frac{CT^2}{H\phi^{n/2}}$, in deep inflationary phase. In particular, we find that $Q=\textrm{const}$ corresponds to $r=2$ and $s=0$  in eq. \eqref{Q-general}, i.e., $Q(\phi,T)=\frac{CT^2}{H}$. 

For consistency, one can check that the $\beta$-function and dissipation coefficient ratio we have just discussed indeed satisfy the relation for quasi-stable radiation energy density given by \eqref{qs-con-c}.
\end{itemize}

\subsection{Trigonometric $\beta$-function}
Let us discuss the  $\beta$-function given by
\begin{equation}
\beta(\phi)=\sqrt{6\lvert\lambda\rvert}\,\tan\left[-\sqrt{\left\vert\frac{3\lambda}{2}\right\vert}\int_{\phi}^{0}(1+Q(\phi'))d\phi'\right]\,.\label{beta-3}
\end{equation}
We find its derivative with respect to $\phi$ as
\begin{widetext}
\begin{equation}
	\beta_{,\phi}=\sqrt{6\lvert\lambda\rvert}\sec^2\left[-\sqrt{\left\vert\frac{3\lambda}{2}\right\vert}\int_{\phi}^{0}(1+Q(\phi'))d\phi'\right]\,\left[\sqrt{\left\vert\frac{3\lambda}{2}\right\vert}(1+Q(\phi))\right]\,.\label{beta-phi-3}
\end{equation}
\end{widetext}
Eq. \eqref{beta-phi-3} tells us that $\beta_{,\phi}\geq 0$ due to $Q(\phi)\geq 0$, hence the $\beta$-function given by eq. \eqref{beta-3} can describe the inflation models with a natural end.
We obtain the superpotential as
\begin{equation}
W(\phi)\simeq\overline{W}_f\cos\left[\sqrt{\left\vert\frac{3\lambda}{2}\right\vert}\int_{0}^{\phi}(1+Q)d\phi'\right]\,,\label{superpotential-sol-3}
\end{equation}
with the normalization constant $\overline{W}_f\equiv W_f\cos^{-1}\left[\sqrt{\left\vert\frac{3\lambda}{2}\right\vert}\int_{0}^{\phi_f}(1+Q)d\phi'\right] $, and the number of e-foldings as
\begin{widetext}
\begin{align}\label{N3}
N=-\int_{\phi_f}^{\phi}\frac{d\phi'}{\beta(\phi')}\simeq\frac{1}{\sqrt{6\vert\lambda\vert}}\int_{\phi}^{\phi_f}\cot\left[-\sqrt{\left\vert\frac{3\lambda}{2}\right\vert}\int_{\phi'}^{0}(1+Q(\phi''))d\phi''\right]d\phi'\,.
\end{align}
The potential is then given by 
\begin{equation}
V(\phi)\simeq\frac{3}{4}\overline{W}_f^2\left\{\cos^2\left[\sqrt{\left\vert\frac{3\lambda}{2}\right\vert}\int_{0}^{\phi}(1+Q(\phi'))d\phi'\right]-\vert\lambda\vert\left(1+\frac{3}{2}Q(\phi')\right)\sin^2\left[\sqrt{\left\vert\frac{3\lambda}{2}\right\vert}\int_{0}^{\phi}(1+Q(\phi'))d\phi'\right]\right\}\,.\label{potential-beta-tan}
\end{equation}
\end{widetext}
We proceed by specifying the dissipation coefficient ratio $Q(\phi)$.
\begin{itemize}
\item $Q=a\,\phi^{n}$, for $n\in \textbf{N}$ and $a\in\textbf{R}$

In this case, eq. \eqref{beta-3} becomes 
\begin{equation}
\beta(\phi)=\sqrt{6\lvert\lambda\rvert}\,\tan\left[\sqrt{\left\vert\frac{3\lambda}{2}\right\vert}\left(\phi+\frac{a\,\phi^{n+1}}{n+1}\right)\right]\label{beta-31}\,.
\end{equation}
Correspondingly, we find the number of e-foldings as
\begin{align}\label{N4}
N\simeq\frac{1}{\sqrt{6\vert\lambda\vert}}\int_{\phi}^{\phi_f}\cot\left[\sqrt{\left\vert\frac{3\lambda}{2}\right\vert}\left(\phi'+\frac{a\,\phi'^{n+1}}{n+1}\right)\right]d\phi'\,.
\end{align}
For the $\beta$-function given by eq. \eqref{beta-31}, the fixed point is placed at $\phi=0$ and  $\phi$ moves away from the fixed point as the universe expands, and finally can reach the value $\phi_f$ at which $\vert\beta(\phi_f)\vert\sqrt{1+Q(\phi_f)}=\sqrt{2}$. Note that $Q(\phi)$ is a monotonically increasing function  of $\phi$ and its value $Q(\phi_f)$ at the end of inflation is controlled by $\lambda$. The smaller $\lambda$ gives rise to the larger $\phi_f$, which in turn corresponds to the larger $\phi$ at horizon crossing for a fixed number of e-foldings (e.g. $N\sim 50$). Thus, depending on $\lambda$, the inflation can occur in different  dissipation regimes. 

For the different values of parameters, the changes in the ratio of radiation energy density to inflaton energy density during inflation are shown together with the corresponding dissipation coefficient ratios in Figs. \ref{rho-Q-2} and \ref{rho-Q-3}.  We see that the inflation can happen in strong dissipative regime  for $\lambda=0.001$ (Fig. \ref{rho-Q-2}), while it can begin in weak dissipative regime and end in strong dissipative regime for $\lambda=0.1$ (Fig.\ref{rho-Q-3}). For both cases, however, the energy density of radiation becomes comparable to that of inflaton field at the end of inflation in spite of the difference in $Q(\phi)$. This is because the amount of energy injected into radiation increases with $\beta^2Q$ (see eq. \eqref{rho_r-rho_phi}) in which  $\beta(\phi)$ decreases and $Q(\phi)$ increases with increasing $\lambda$.

	\begin{figure}[htp!]
		\centering
		\begin{subfigure}{0.23\textwidth}
			\includegraphics[width=1\linewidth,height=3cm]{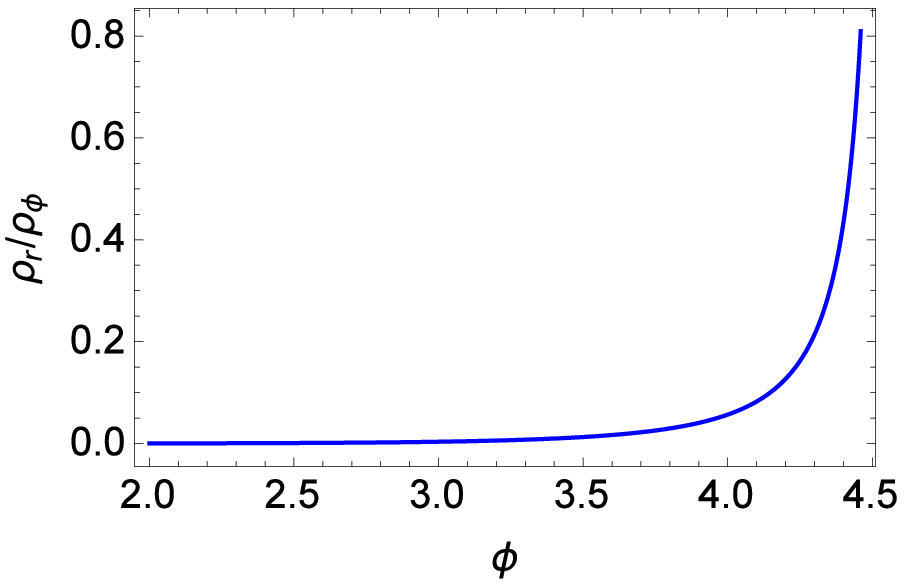}
		\end{subfigure}
		\begin{subfigure}{0.23\textwidth}
			\includegraphics[width=1\linewidth,height=3cm]{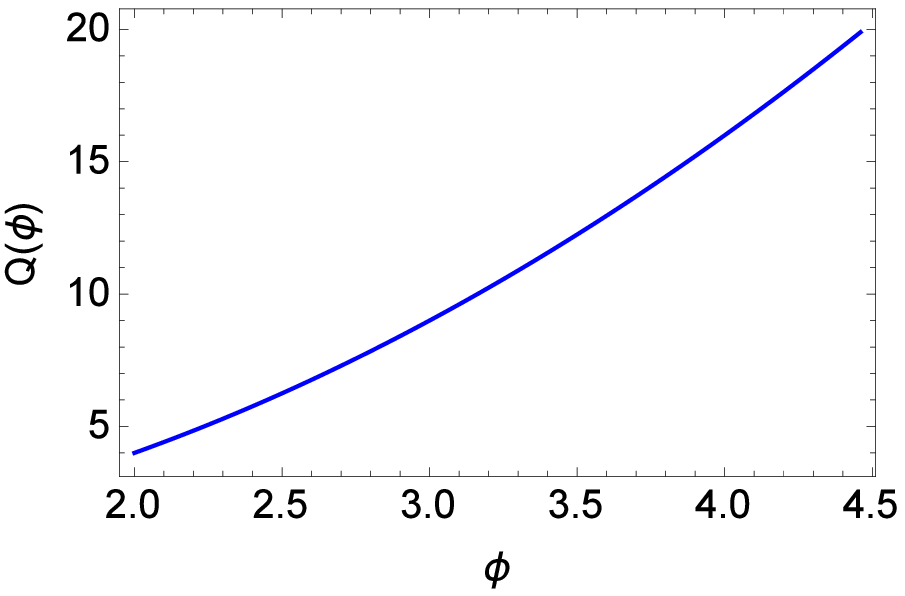}
		\end{subfigure}
		\begin{subfigure}{0.23\textwidth}
			\includegraphics[width=1\linewidth,height=3cm]{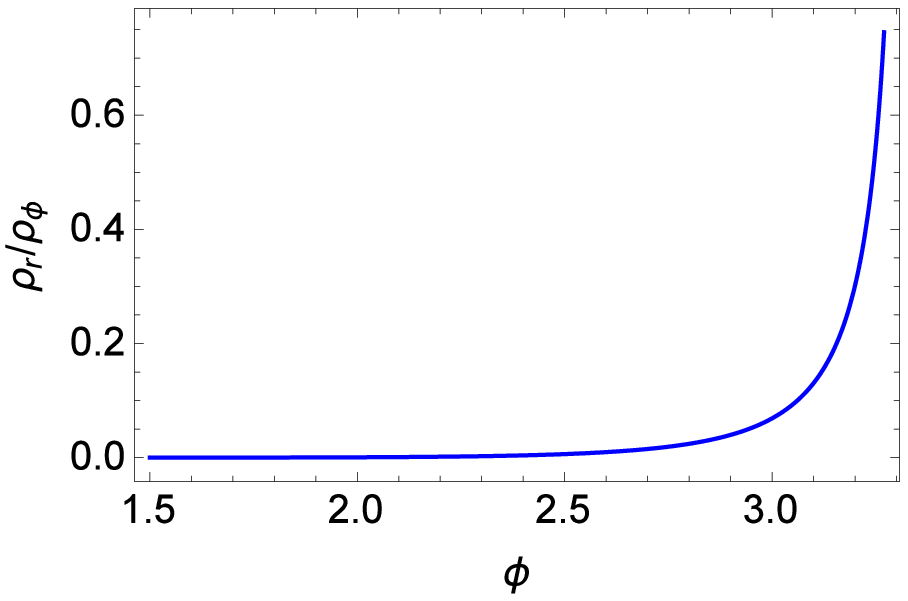}
		\end{subfigure}
		\begin{subfigure}{0.23\textwidth}
			\includegraphics[width=1\linewidth,height=3cm]{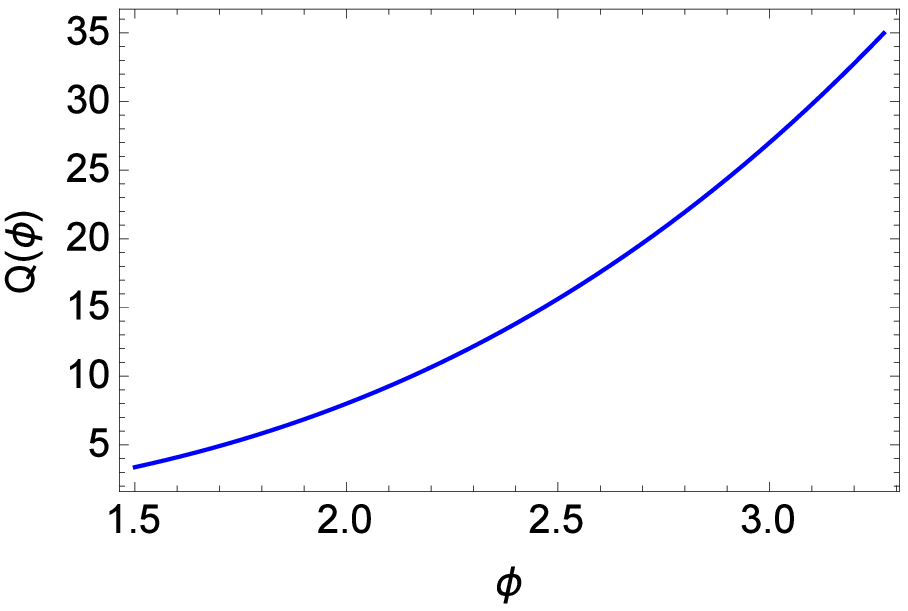}
		\end{subfigure}
		\caption{The evolution of $\rho_r/\rho_\phi$ and $Q(\phi)=\phi^{n}$ for  $\beta(\phi)$  given by eq. \eqref{beta-31} with  $\lambda=0.001$ during inflation. We have chosen $a=1$, $n=2$ (top) and $a=1$, $n=3$ (bottom). The range of $\phi$ in each plot was chosen such that the number of e-foldings satisfies $N\sim 50$.}\label{rho-Q-2}
	\end{figure}  
	\begin{figure}[htp!]
	\centering
	\begin{subfigure}{0.23\textwidth}
		\includegraphics[width=1\linewidth,height=3cm]{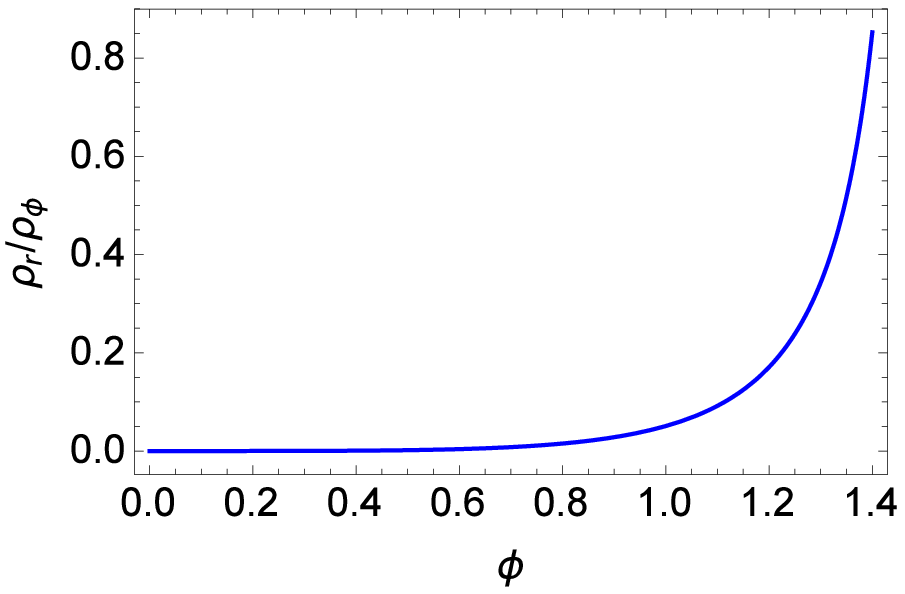}
	\end{subfigure}
	\begin{subfigure}{0.23\textwidth}
		\includegraphics[width=1\linewidth,height=3cm]{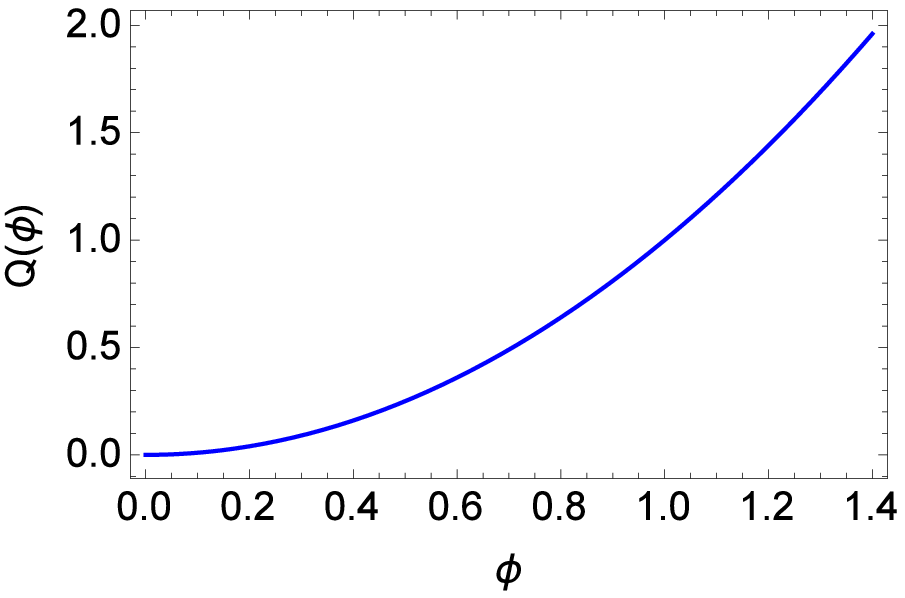}
	\end{subfigure}
	\begin{subfigure}{0.23\textwidth}
		\includegraphics[width=1\linewidth,height=3cm]{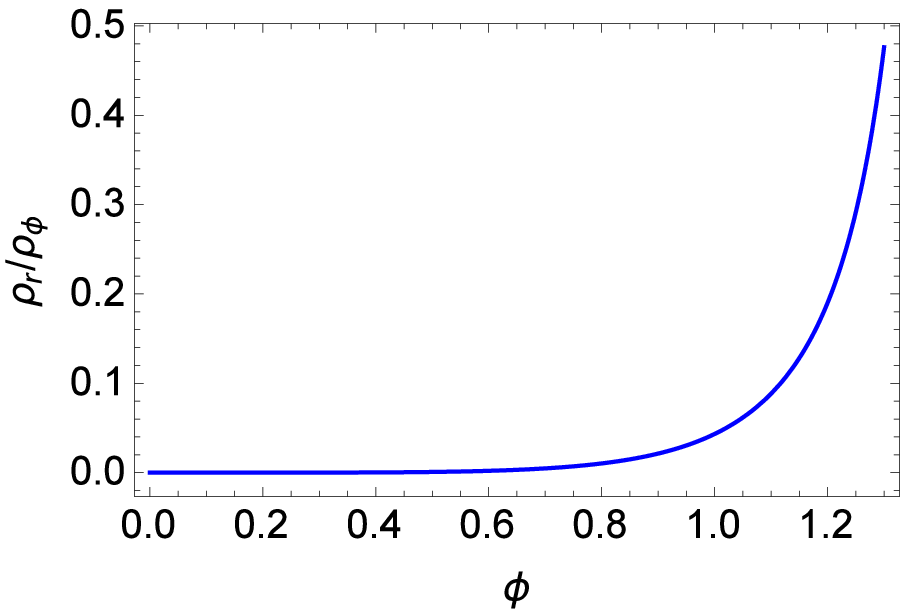}
	\end{subfigure}
	\begin{subfigure}{0.23\textwidth}
		\includegraphics[width=1\linewidth,height=3cm]{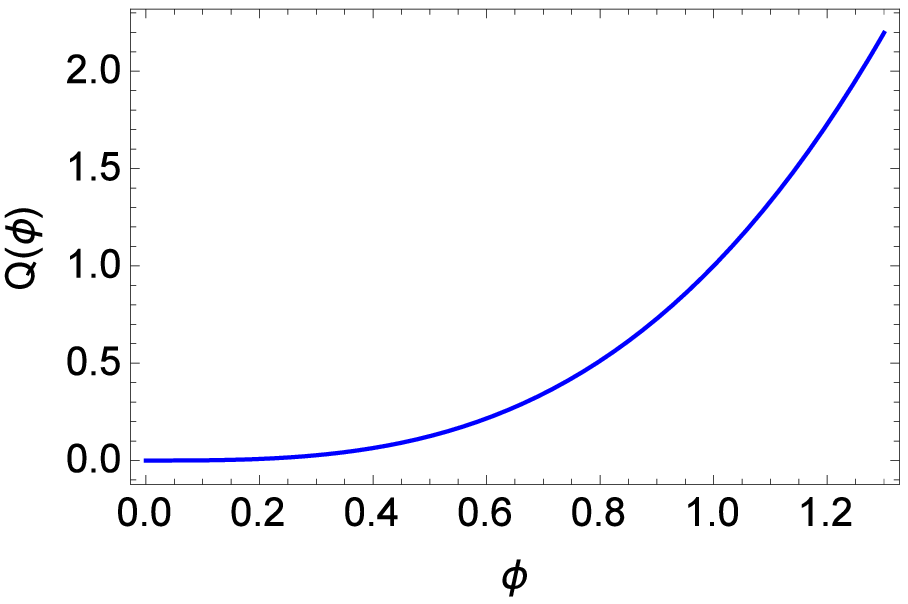}
	\end{subfigure}
	\caption{The evolution of $\rho_r/\rho_\phi$ and $Q(\phi)=\phi^{n}$ for $\beta(\phi)$ given by eq. \eqref{beta-31} with  $\lambda=0.1$ during inflation. We have chosen $a=1$, $n=2$ (top) and $a=1$, $n=3$ (bottom). The range of $\phi$ in each plot was chosen such that the number of e-foldings satisfies $N\sim 50$. }\label{rho-Q-3}
\end{figure}

The temperature of the radiation bath as a function of $\phi$ can also be found  by using eqs. \eqref{betaQ-rho-r}, \eqref{rho-r-T} together with eq. \eqref{superpotential-sol-3} as
\begin{equation}
	T(\phi)\simeq\sqrt{\left\vert\overline{W}_f\sqrt{6a\vert\lambda\vert\phi^n}\,\sin\left[\sqrt{\left\vert\frac{3\lambda}{2}\right\vert}\left(\phi+\frac{a\,\phi^{n+1}}{n+1}\right)\right]\right\vert}\,.\label{T-phi-2}
\end{equation}  
Using eqs. \eqref{superpotential-sol-3} and \eqref{T-phi-2}, we find that in deep inflationary phase, the dissipation coefficient ratio $Q=a\phi^n$ corresponds to the general parameterization in eq. \eqref{Q-general} in which the parameters $r,s$ satisfy $\frac{r/2-s}{1-r/4}=n$. This suggests that there is a degeneracy in $Q(T,\phi)$ corresponding to single $Q(\phi)$, as discussed in Ref. \cite{berera}.

\item $Q=a\,\phi^{-n}$, for $n\in \textbf{N}$ and $a\in\textbf{R}$

The $\beta$-function is
\begin{equation}
\beta(\phi)=\sqrt{6\lvert\lambda\rvert}\,\tan\left[\sqrt{\left\vert\frac{3\lambda}{2}\right\vert}\left(\phi+\frac{a\,\phi^{1-n}}{1-n}\right)\right]\label{beta-32}\,,
\end{equation}
which gives the number of e-foldings
\begin{align}\label{N5}
N\simeq\frac{1}{\sqrt{6\vert\lambda\vert}}\int_{\phi}^{\phi_f}\cot\left[\sqrt{\left\vert\frac{3\lambda}{2}\right\vert}\left(\phi'+\frac{a\,\phi'^{1-n}}{1-n}\right)\right]d\phi'\,.
\end{align}
From eq. \eqref{beta-32}, we see that $\beta(\phi)$ has a fixed point at $\phi=\left(\frac{a}{n-1}\right)^{1/n}$.

Inflation can be realized in both cases where $\phi$ moves to  right (i.e. $\phi$ increases) or to  left (i.e. $\phi$ decreases) from the fixed point. Since $Q(\phi)$ is monotonically decreasing with increasing $\phi$, it behaves conversely in two cases. Hence, depending on the direction in which $\phi$ moves from the fixed point, inflation can occur in weak dissipative regime or strong dissipative regime. This is illustrated in Fig. \ref{rho-Q-4} including the corresponding ratios of radiation energy density to inflaton energy density. We see from Fig. \ref{rho-Q-4} that  $\rho_r/\rho_\phi$ in both cases have the same behavior in spite of totally different behavior of $Q(\phi)$. This is again due to the fact that $\rho_r/\rho_\phi$ is controlled by $\beta^2Q$ rather than $Q$ only.
	\begin{figure}[htp!]
	\centering
	\begin{subfigure}{0.23\textwidth}
		\includegraphics[width=1\linewidth,height=3cm]{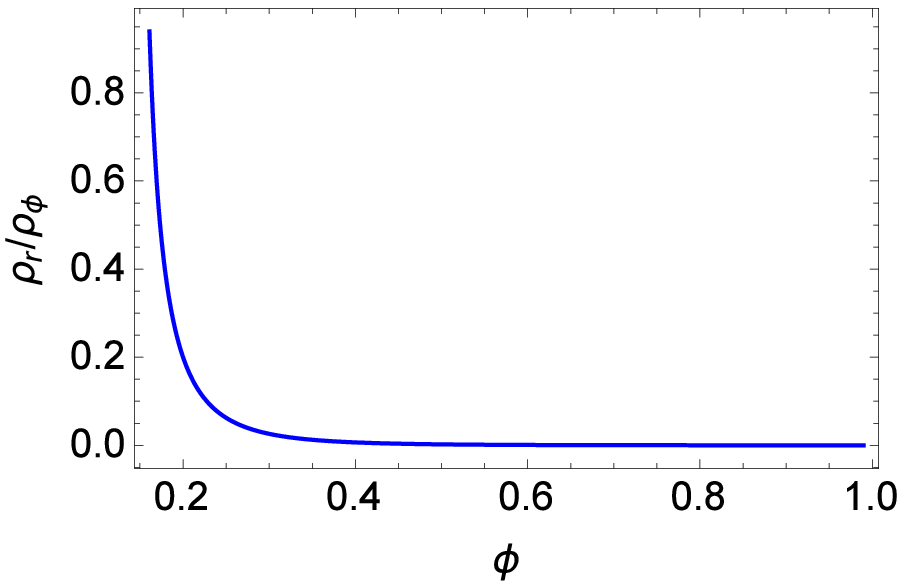}
	\end{subfigure}
	\begin{subfigure}{0.23\textwidth}
		\includegraphics[width=1\linewidth,height=3cm]{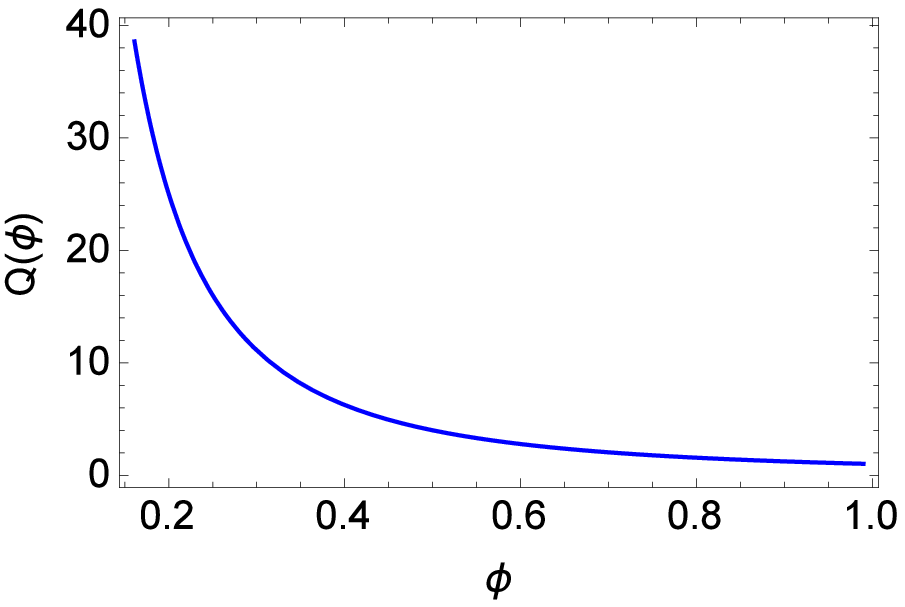}
	\end{subfigure}
	\begin{subfigure}{0.23\textwidth}
		\includegraphics[width=1\linewidth,height=3cm]{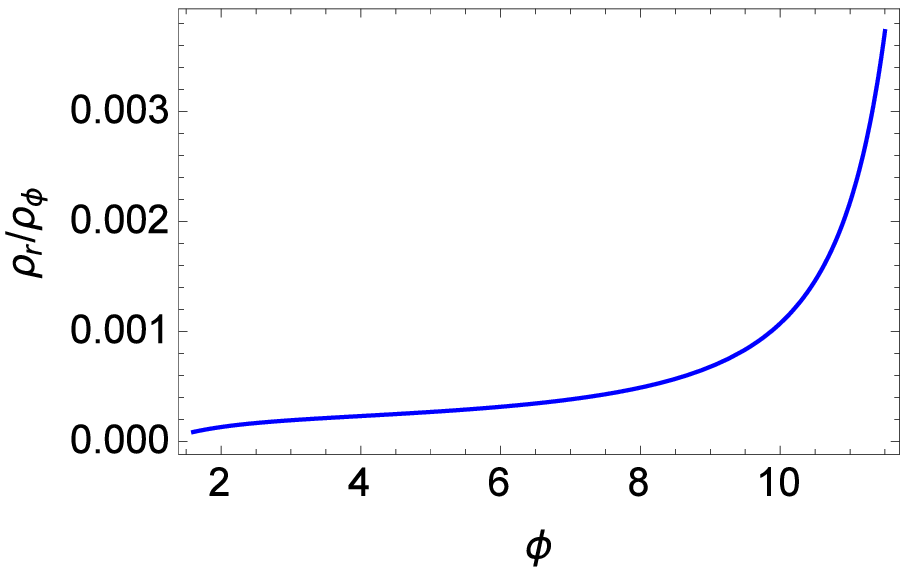}
	\end{subfigure}
	\begin{subfigure}{0.23\textwidth}
		\includegraphics[width=1\linewidth,height=3cm]{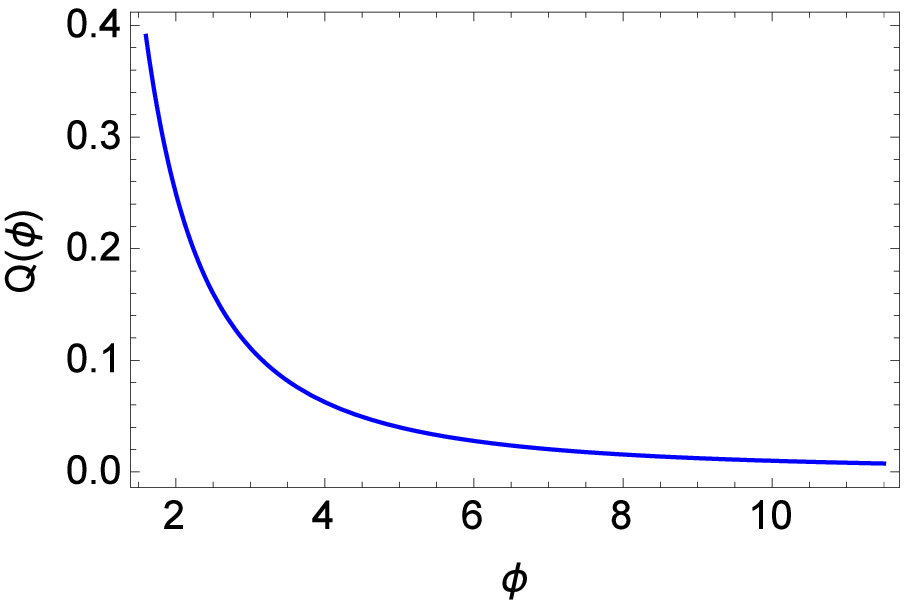}
	\end{subfigure}
	\caption{The evolution of $\rho_r/\rho_\phi$ and $Q(\phi)=\phi^{-n}$ for $\beta(\phi)$ given by eq. \eqref{beta-32} for  $\lambda=0.01$, $a=1$ and $n=2$. The range of $\phi$ in each plot was chosen such that the number of e-foldings satisfies $N\sim 50$. The top plots correspond to the case in which $\phi$ moves from the fixed point to left so that the inflation can occur in strong dissipative regime and the bottom plots correspond to the case in which $\phi$ moves to right so that the inflation can occur in weak dissipative regime.}\label{rho-Q-4}
\end{figure}  
The temperature of the radiation bath is obtained as
\begin{equation}
		T(\phi)\simeq\sqrt{\left\vert\overline{W}_f\sqrt{6a\vert\lambda\vert\phi^{-n}}\,\sin\left[\sqrt{\left\vert\frac{3\lambda}{2}\right\vert}\left(\phi+\frac{a\,\phi^{1-n}}{1-n}\right)\right]\right\vert}\,.\label{T-phi-3}
\end{equation}
\end{itemize}
Soon after $\phi$ starts to move from the fixed point to left, one observes that the second term becomes dominant in the argument of sine function in eq. \eqref{T-phi-3}, i.e., $\vert\phi\vert\ll \left\vert\frac{a\phi^{1-n}}{1-n}\right\vert$. This together with eqs. \eqref{T-phi-3} and \eqref{superpotential-sol-3} allows us to identify $Q=a\phi^{-n}$ with the general parameterization given by eq. \eqref{Q-general} with $r=2, \,s=1$, i.e., $Q(T,\phi)=\frac{CT^2}{H\phi}$. 
On the other hand, when $\phi$ moves from the fixed point to right, we find that near the end of inflation $Q=a\phi^{-n}$ corresponds to eq. \eqref{Q-general} with $r=2,\, s=n+1$, i.e., $Q(T,\phi)=\frac{CT^2}{H\phi^{n+1}}$. 

We point out that the $\beta$-functions and dissipation coefficient ratios we have  discussed in this section satisfy the relation for quasi-stable radiation energy density given by \eqref{qs-con-c}.\\

The investigation of the last solution to eq. \eqref{const-roll-2} is so similar to the previous one that we will not do it here. Notice that the last solution  to eq. \eqref{const-roll-2} is not equivalent to the second solution by the redefinition of $\phi\to \frac{\pi}{\sqrt{6\vert\lambda\vert}}-\phi$ as in the case of cold inflation \cite{Francesco}, due to the presence of non-trivial $Q(\phi)$.  We close this section with summarizing the $\beta$-functions and the dissipation coefficient ratios that we have discussed, which can give constant-roll warm inflation (see Table. \ref{tab-1}).
\begin{table*}[htp!]
	\centering
	\caption{Some possible $\beta$-functions and  dissipation coefficient ratios for constant-roll warm inflation }\label{tab-1}
	\begin{tabular}{c|c|c}
		\hline
		$\beta(\phi)$ & $Q(\phi)$ & $Q(T, \phi)$\\\hline
		$\beta(\phi)=\sqrt{6\lambda}\coth\left[-\sqrt{\frac{3\lambda}{2}}\left(\phi+\frac{a\,\phi^{1-n}}{1-n}\right)\right]$ & $a\phi^{-n}$ & $\frac{CT^2}{H\phi^{n/2}}$ (in deep inflationary phase)\\\hline
		$\beta(\phi)=\sqrt{6\lvert\lambda\rvert}\,\tan\left[\sqrt{\left\vert\frac{3\lambda}{2}\right\vert}\left(\phi+\frac{a\,\phi^{n+1}}{n+1}\right)\right]$ & $a\phi^n$ & $\frac{CT^r}{H\phi^s}$ with $\frac{r/2-s}{1-r/4}=n$  (in deep inflationary phase)\\\hline
$\beta(\phi)=\sqrt{6\lvert\lambda\rvert}\,\tan\left[\sqrt{\left\vert\frac{3\lambda}{2}\right\vert}\left(\phi+\frac{a\,\phi^{1-n}}{1-n}\right)\right]$ & $a\phi^{-n}$ & $\frac{CT^2}{H\phi}$ or $\frac{CT^2}{H\phi^{n+1}}$ (near the end of inflation)\\\hline		
		
	\end{tabular}
\end{table*}

\section{Conclusions}\label{sec-3}
We have proposed a new approach to constant-roll warm inflation and investigated its realization by adopting $\beta$-function approach. 

It has been shown that the  constant-roll warm inflationary condition can  be translated to a simple first order differential equation of $\beta(\phi)$ which is solvable analytically as in cold constant-roll inflation, in the presence of non-trivial dissipation coefficient ratio $Q(\phi)$. Given that it would be impossible to determine the possible forms of the effective potential of inflaton field from the constant-roll warm inflationary condition, without knowing the explicit form of $Q(\phi)$, this result shows one advantage of using $\beta$-function formalism in the study of constant-roll warm inflation.

Based on the argument that $\beta$-functions must have an ``infrared" fixed point, or more generally, to be an increasing function of $\phi$, in order to give a natural end of inflation, we have found three possible types of $\beta$-functions for constant-roll warm inflation. For two types of $\beta$-functions (hyperbolic tangent and tangent), we have considered two special cases of $Q(\phi)$ and investigated whether or not they can provide a physically plausible inflation model. We have found that for some  dissipation coefficient ratios, the quantity $\beta(\phi)\sqrt{1+Q(\phi)}$ has the asymptotic behavior different from that in cold inflation in deep inflationary phase and becomes inappropriate for inflation. 

It has been demonstrated that constant-roll warm inflation may occur completely in weak or strong dissipative regime depending on the direction in which $\phi$ moves from the fixed point as well as on the model parameters such as $a$ in $Q(\phi)$ and $\lambda$ in $\beta(\phi)$.  

We have finally discussed the correspondence between two different parameterizations of the dissipation coefficient ratio, namely $Q(\phi)$ and $Q(T,\phi)$, and illustrated the degeneracy in the parameterization of $Q(T,\phi)$ for single $Q(\phi)$. This supports the idea proposed in Ref. \cite{berera} that specifying  $Q(\phi)$ would be more general than  specifying $Q(T,\phi)$ in the study of universality classes of warm inflation models.

We have not discussed the predictions of constant-roll warm inflation for CMB observables, namely, the scalar spectral index $n_s$ and tensor-to-scalar ratio $r$, which requires the knowledge of radiation fluctuations coupled to inflaton field \cite{future-1,future-2} and rather involved and careful treatment due to the breakdown of slow-roll condition. We leave this for the future work.
Finally, as an alternative to our generalization of constant-roll condition, it would also be interesting to consider the case in which $\ddot{\phi}/3H\dot{\phi}=\textrm{const}$ but $\ddot{\phi}/3H(1+Q)\dot{\phi}\neq \textrm{const}$ for non-trivial $Q$.

\section*{Acknowledgements}
We thank Jin U Kang and Ok Song An for very helpful discussions and comments on the manuscript.

\bibliographystyle{apsrev}

\end{document}